# LARGE DIFFUSE HALOS IN TIME-DEPENDENT SPACE-CHARGE POTENTIALS WITH COLORED NOISE*

C. L. Bohn[#], Northern Illinois University, DeKalb, IL 60115 and Fermilab, Batavia, IL 60510, USA
I. V. Sideris, Northern Illinois University, DeKalb, IL 60115, USA

*Abstract*

We explore the potential impact of colored noise on space-charge-induced halo formation. By coupling particle orbits to parametric resonance, colored noise due to space-charge fluctuations and/or imperfections in the beamline can eject particles to much larger amplitudes than would be inferred from parametric resonance alone.

## INTRODUCTION

The pursuit of high-current light-ion accelerators such as are needed to drive high-intensity spallation neutron sources has brought the realization that, albeit necessary, the ability to control root-mean-square beam properties is not sufficient. Perhaps the most prominent example concerns beam halos, *i.e.*, particles that reach large orbital amplitudes due to time dependence in the space-charge potential. The concern is that impingement of beam particles on the beamline could generate radioactivation that would preclude routine, hands-on maintenance. Just a tiny impingement (~1 nA/m/GeV) is troublesome. For a 100 mA, 1 GeV light-ion beam, this criterion translates to just 1 in $10^8$ particles lost per meter [1].

Early efforts to identify the fundamental mechanisms of halo formation centered on the use of a 'particle-core' model [2,3]. The basic recognition was that if a uniform-density core is made to pulsate, particles that initially lie outside the core and that resonate with its pulsations can reach large amplitudes and form a 'halo'. This led to the identification of parametric resonance as the essential mechanism of halo formation.

The key feature of parametric resonance in the context of the particle-core model is a hard upper bound to the amplitude that a halo particle can reach [4]. The particle's orbital frequency is a function of its amplitude. At large amplitude the particle falls out of resonance with the core, thereby preventing the amplitude from growing further. This lends hope to the possibility of computing an aperture requirement for beamline components; smaller aperture requirements are preferred in that, for example, they mitigate concerns about shunt impedance in the accelerating cavities.

An important question is whether there is any physics not included in the particle-core model that could significantly influence the maximum particle amplitude. Herein we show that the answer is yes; *the presence of colored noise can potentially boost a statistically rare particle to much larger amplitudes by continually kicking it back into phase with the core oscillation.*

*Work supported by DoEd Grant G1A62056.
[#]clbohn@fnal.gov

## METHODOLOGY

Following the ground-breaking work that introduced the particle-core model [2,3], we consider radial orbits governed by the dimensionless equation of motion

$$\ddot{x} + x[(1 - x^{-2})\Theta(|x|-1) + 2(M-1)\cos\omega t\, \Theta(1-|x|)] = 0, \quad (1)$$

where $x$ is normalized to the matched core radius $a$, time is multiplied by the external focusing frequency $\Omega$, $\omega$ is the core-oscillation frequency, $M=a(0)/a$ is the mismatch, i.e., the ratio of initial-to-matched core radii, and $\Theta(u)$ is the Heaviside step function. The first and second terms in square brackets govern the motion of the particle when it is outside and inside the core, respectively.

We add to Eq. (1) fluctuations in the form of Gaussian colored noise such that $\omega \to \omega(t) = \omega_0 + \delta\omega(t)$, with

$$\langle \delta\omega(t) \rangle = 0 \quad \text{and} \quad \langle \delta\omega(t)\delta\omega(t_1) \rangle \propto \exp(-|t-t_1|/t_c), \quad (2)$$

in which $t_c$ is the autocorrelation time. Upon generating a colored-noise signal, we calculate $\langle |\delta\omega| \rangle$ which becomes a measure of the noise strength. Colored noise is a realistic phenomenon that will arise self-consistently from charge-density variations and from irregularities in the beamline; in the former case the autocorrelation time could be short, say of the order of a plasma period, and in the latter case it could be long, say several betatron (orbital) periods [5].

We fix the mismatch parameter at $M=1.5$, and the initial conditions of the orbit at $x(0)=1.2$, $\dot{x}(0)=0$. The orbit is computed from Eq. (1) first without, then with, the noise of Eq. (2) using a variable-time-step integrator. For zero colored noise, the maximum amplitude $|x_{max}|$ depends on the core-oscillation frequency $\omega_0$; Fig. 1 shows this dependence. The particle can reach relatively large amplitudes for a wide range of frequencies $\omega_0$. We present results for $\omega_0$=3.53 because the associated power spectra of the orbits are especially clear; different choices of $\omega_0$ do not change the essential findings.

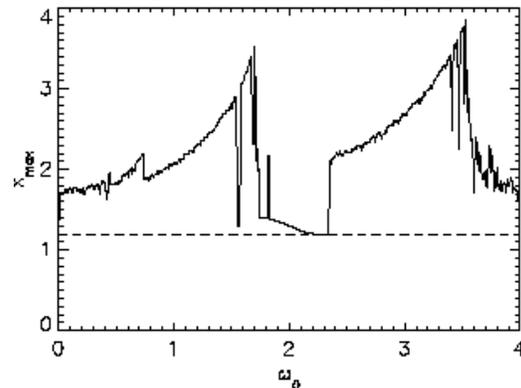

Fig. 1. Maximum orbital amplitude $|x_{max}|$ vs. frequency $\omega_0$ with zero noise. The horizontal line is at $x(0)=1.2$.

We consider the colored noise to be a random process, which means each individual particle will see a different manifestation of the noise. Thus, in an 'axisymmetric' model, such as that of Eq. (1), each particle initially occupying a thin annulus centered at radius $x(0)$ will experience noise that differs from that seen by each of the other particles. Accordingly, we adopted a 'survey strategy'. For the given initial conditions and a specific choice of noise parameters, we sequentially computed 10,000 orbits, each experiencing its own random manifestation of the colored noise, and we catalogued the maximum amplitudes of these orbits. As is shown and explained below, the colored noise will keep a statistically rare particle in phase with the core oscillation such that this particle reaches much larger amplitudes than without colored noise. Moreover, as time progresses, its maximum amplitude will grow indefinitely. We shall demonstrate that there is in principle no hard upper bound to the halo amplitude in the presence of colored noise.

## RESULTS

### Without Colored Noise

The case of zero noise constitutes the baseline against which to assess the influence of colored noise. The corresponding orbit $x(t)$ and its power spectrum appear in Fig. 2. The orbit, which we integrated to high precision, reaches a maximum radius $x_{max}=3.9$; it goes no further regardless how long the orbit is integrated. Parametric resonance indeed establishes a hard upper bound to the orbital amplitude in the absence of noise.

The power spectrum, however, is complicated, and stays complicated. As Fig. 2 suggests, there is interplay between harmonics of the external frequency $\Omega$ and core-oscillation frequency $\omega_0$ (since time is measured in the unit of the inverse external frequency, $\Omega = 1$ here). The interplay results in the formation of not only several prominent spectral lines, but also continua, signifying that the orbit is chaotic.

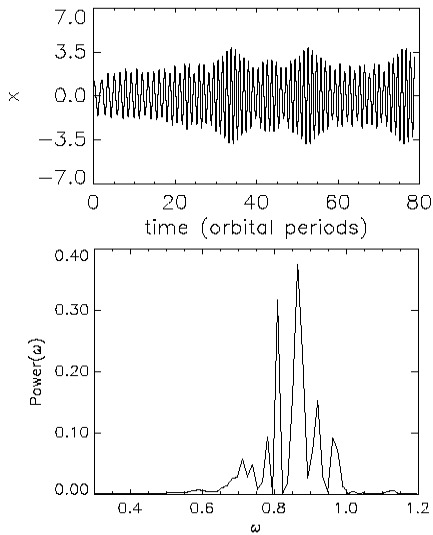

Fig. 2. (top) Orbit without noise; (bottom) power spectrum (Fourier transform of the orbit).

### With Colored Noise

The influence of noise on halo formation depends on its strength $\langle|\delta\omega|\rangle$ and its autocorrelation time. For two choices of autocorrelation time, $t_c = 1\tau$ and $t_c = 12\tau$, with $\tau$ denoting the representative orbital period of the particle, we investigated a broad range of strengths, specifically $10^{-5} \leq \langle|\delta\omega|\rangle \leq 1$, with the goal of ascertaining to what extent the results may be regarded as generic. All of the results presented herein derive from sequential computations of 10,000 orbits with noise randomly generated for each orbit separately.

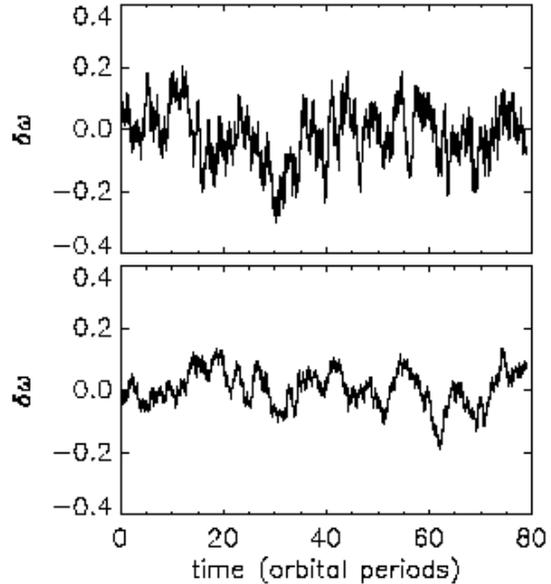

Fig. 3. Example manifestations of colored noise along an orbit for $\langle|\delta\omega|\rangle=0.09$ and $t_c=1.5\tau$ (top), $12\tau$ (bottom).

Manifestations of colored noise that a particle might see are illustrated in Fig. 3, which is provided as an aid toward conceptualizing the physical meaning of the noise parameters. Shown there are manifestations of noise for $t_c= 1.5\tau$ and $12\tau$ for a fixed range $\langle|\delta\omega|\rangle=0.09$.

Next, we consider the one particle out of the sample of 10,000 that reaches to the largest amplitude $|x_{max}|$ during the integration time of $80\tau$, which is representative of the length of a 1 GeV proton linac. Results for fixed $t_c = 12\tau$ and $\langle|\delta\omega|\rangle=0.09$ are provided in Fig. 4, which shows this special orbit. Figure 4 also shows the distribution of $|x_{max}|$ for the entire 10,000-orbit sample. Notice that for this case $|x_{max}|$ averaged over the sample is less than 3.9, the value pertaining to the bare parametric resonance. This means the noise kicks 'typical' particles out of phase with the core oscillation while it simultaneously kicks rare particles into phase. Also plotted is $|x_{max}|$ versus $\langle|\delta\omega|\rangle$, from which one sees that even much weaker noise can suffice to eject rare particles to larger amplitudes relative to the parametric resonance alone. This is due to the chaoticity of the orbits in that their inherent continuum of frequencies enhances the coupling of the noise to the orbit even when $\langle|\delta\omega|\rangle$ is small. The power spectrum of these rare orbits eventually becomes sharply peaked at the particle's orbital frequency. Interestingly, we find for $t_c = 1\tau$ that there is no substantial difference in the results; they are qualitatively the same.

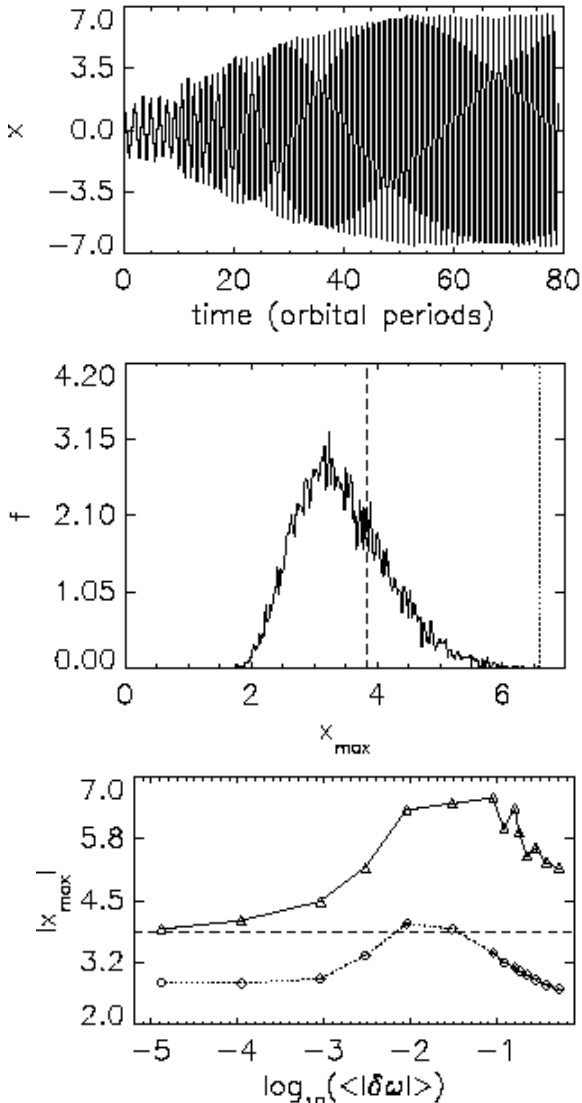

Fig. 4. (top) Orbit of the one particle reaching the largest $|x_{max}|$ given noise with $\langle|\delta\omega|\rangle$=0.09, $t_c = 12\tau$. (middle) Distribution of $|x_{max}|$ of all 10,000 particles in the sample; the lines show largest $|x_{max}|$ with no noise (dashed) and with this noise (dotted). (bottom) Largest $|x_{max}|$ reached by any particle vs. $\langle|\delta\omega|\rangle$ with $t_c = 12\tau$ (triangles); mean $|x_{max}|$ of the 10,000-particle sample (diamonds).

If the integration time is extended indefinitely, then there are orbits that continue to grow to seemingly unlimited amplitude. The orbit of one such particle appears in Fig. 5. This long-time orbit exemplifies that *there is in principle no upper bound to the halo amplitude in the presence of colored noise*.

## DISCUSSION

The influence of colored noise on particle orbits, and in particular its role in generating large diffuse halos in time-dependent potentials, appears to be generic, i.e., not restricted to the simple particle-core model of Eq. (1). For example, we did the same numerical experiments using a model of a beam bunch in which the unperturbed potential is a spherically symmetric configuration in thermal equilibrium [6] and the perturbation has prolate spheroidal symmetry with a single driving frequency. In this potential the vast majority of the orbits quickly became chaotic and coupled to the perturbation. The net effect as concerns halo production was qualitatively the same as described above, though the dynamical details were richer, reflecting the more complicated potential. The collection of findings suggests that formation of large diffuse halos is not particularly sensitive to details in either the governing potential or the noise.

It remains, of course, to explore the extent to which this phenomenology applies in a real machine. Doing so will involve simulations of beams in real beamlines. One possibly fruitful approach is to extract the smooth, time-dependent potential from the simulations and then pursue a statistical analysis in parallel to what we have done here. Alternatively, the colored noise may be built into the simulation itself. A realistic manifestation of the colored noise would need to reflect the machine design, i.e., by incorporating imperfections in the fields and hardware alignment, and/or by including details of the evolving space-charge potential such as a sufficiently detailed mode spectrum. Of course, as the beam is accelerated and becomes relativistic, space charge will become decreasingly important, and growth of the halo will thereby be mitigated.

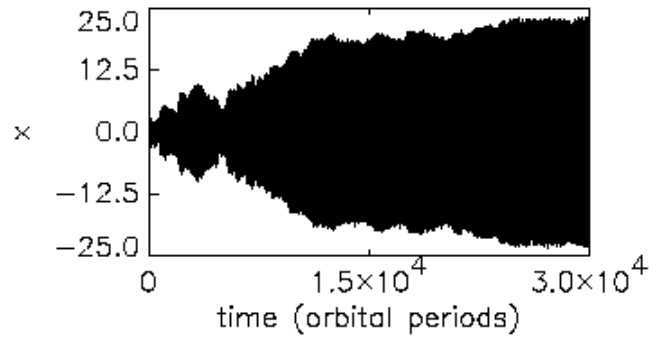

Fig. 5. Long-time evolution of a large-amplitude orbit given noise with $\langle|\delta\omega|\rangle$=0.09, $t_c = 12\tau$.